\documentclass[10pt,A4paper,amsmath,amssymb,aps,etoolbox,floatfix,
longbibliography,nofootinbib,noshowpacs,onecolumn,preprintnumbers,reprint,
showkeys,showpacs,superscript address]{revtex4}
\newcommand{\bea}{\begin{eqnarray}}
\newcommand{\beq}{\begin{equation}}

\newcommand{\eea}{\end{eqnarray}}
\newcommand{\eeq}{\end{equation}}
\newcommand{\ds}{\displaystyle}

\usepackage{amsmath}
\usepackage{amsfonts}
\usepackage{amssymb}
\usepackage{bm}       % bold math
\usepackage{caption}
\usepackage{comment}
\usepackage[usenames,dvipsnames]{color}
\usepackage{dcolumn}  % Align table columns on decimal point
\usepackage{epsfig}
\usepackage{float}
\usepackage{graphics}
\usepackage{graphicx} % Include figure files
\usepackage{natbib}
\usepackage{capt-of} % to put figure and table in one float, with proper captions
\usepackage{placeins} % for \FloatBarrier
\usepackage{psfrag}
\usepackage{times}
\usepackage[varg]{txfonts}
\usepackage{xcolor}
\usepackage{xspace}

\begin{document}

\title{Investigating dark energy by electromagnetic frequency shifts II: the Pantheon+ sample}

\author{Giuseppe Sarracino}

\affiliation{{\mbox Dipartimento di Fisica E. Pancini, Universit\`{a} degli Studi di Napoli Federico II (UNINA)}\\ 
{\mbox Complesso Universitario Monte S. Angelo, Via Cinthia 9 Edificio G, 80126 Naples, Italy}}

\affiliation{{\mbox  Sezione di Napoli, Istituto Nazionale di Fisica Nucleare (INFN)}\\
{\mbox Complesso Universitario Monte S. Angelo, Via Cinthia 9 Edificio G, 80126 Naples, Italy}}

\author{Alessandro D.A.M. Spallicci}
\affiliation{\mbox{Institut Denis Poisson (IDP) UMR 7013}\\
\mbox{Centre National de la Recherche Scientifique (CNRS)}\\
\mbox{Universit\'e d'Orl\'eans (UO) et Universit\'e de Tours (UT)}\\
\mbox{Parc de Grandmont, 37200 Tours, France}}

\affiliation{\mbox{Laboratoire de Physique et Chimie de l'Environnement et de l'Espace (LPC2E) UMR 7328}\\
\mbox{Centre National de la Recherche Scientifique (CNRS)}\\
\mbox{Centre National d'\'Etudes Spatiales (CNES)}\\
\mbox{Universit\'e d'Orl\'eans (UO)}\\
\mbox {3A Avenue de la Recherche Scientifique, 45071 Orl\'eans, France}}

\affiliation{\mbox{UFR Sciences et Techniques, 
Universit\'e d’Orl\'eans (UO)}\\
\mbox{Rue de Chartres, 45100 Orl\'{e}ans, France}}

\affiliation{\mbox{Observatoire des Sciences de l'Univers en region Centre (OSUC) UMS 3116,%}\mbox{
Universit\'e d'Orl\'eans (UO)}\\
\mbox{1A rue de la F\'{e}rollerie, 45071 Orl\'{e}ans, France}}

\author{Salvatore Capozziello}

\affiliation{{\mbox Dipartimento di Fisica E. Pancini, Universit\`{a} degli Studi di Napoli Federico II (UNINA)}\\ 
{\mbox Complesso Universitario Monte S. Angelo, Via Cinthia 9 Edificio G, 80126 Napoli, Italy}}

\affiliation{{\mbox  Sezione di Napoli, Istituto Nazionale di Fisica Nucleare (INFN)}\\
{\mbox Complesso Universitario Monte S. Angelo, Via Cinthia 9 Edificio G, 80126 Napoli, Italy}}

\affiliation{{\mbox Scuola Superiore Meridionale}\\ {\mbox Largo San Marcellino 10, 80138 Naples, Italy} \\
giuseppe.sarracino@unina.it; alessandro.spallicci@cnrs-orleans.fr; salvatore.capozziello@unina.it}

\date{December 2022}

\begin{abstract}
Following results presented in  Spallicci et al. (Eur Phys J Plus 137, 2022) by the same authors, we  investigate  the observed red shift $z$, working under the hypothesis that it might be composed by the expansion red shift $z_{\rm C}$ and an additional frequency shift $z_{\rm S}$, towards the red or the blue, due to Extended Theories of Electromagnetism (ETE). We have tested this prediction considering the novel Pantheon+ Catalogue, composed by 1701 light curves collected by 1550 SNe Ia, and 16 BAO data, for different cosmological models characterised by the absence of a dark energy component. In particular, we shall derive which values of $z_{\rm S}$  match the observations, comparing the new results with the ones obtained considering the older Pantheon Catalogue. We find interesting differences in the resulting $z_{\rm S}$ distributions, highlighted in the text. Later, we also add a discussion regarding Extended Theories of Gravity and how to incorporate them in our methodology.
\end{abstract}


\maketitle

\section{Introduction}
The Lambda Cold Dark Matter ($\Lambda$CDM) model is, up to now, the most successful cosmological model, able to fit the majority of the cosmological observations with a limited number (6) of free parameters.  According to this model, the Universe is filled up for its majority by unknown fundamental constituents, namely dark matter and dark energy. Indeed, the former would constitute around $25\%$ while the latter around $70\%$ of the total density of the Universe at our epoch \cite{riess-etal-1998,riess-etal-2004,perlmutter-etal-1999,bahcall-etal-1999,spergel-etal-2003,schimd-etal-2006,mcdonald-etal-2006,bamba-etal-2012,joyce-etal-2015,salucci-etal-2021}. Dark matter was firstly introduced in order to fill the "missing matter" found in galaxy clusters which should have been detected according to the virial theorem \cite{zwicky-1933}. Later, its contribution has been proven fundamental also for the galactic rotation curves. Indeed, an excess in the velocity as a function of radial distance to the centre has been observed, which without the introduction of dark matter would not be well fitted by Newtonian computations (that should work at these scales \cite{Babcock1939, Rubin1970}). Dark energy, instead, has been reintroduced (after the first formulation of a cosmological constant due to Einstein for keeping a stationary universe model) to explain the discrepancy between the inferred luminosity distance of Supernovae type Ia (SNe Ia), considered to be well established standard candles \cite{tripp-1998}, and the associated red shift \cite{riess-etal-1998,perlmutter-etal-1999}, for a universe without dark energy. 

Despite the ability to fit the observations via these constituents, the main issue with this approach is the lack of direct observations, despite the investments by the scientific community, for both these quantities, even with very high precision experiments \cite{pigozzo-etal-2011,lopezcorredoira-2017, Xenon2017, Workman-etal-2022}. Also, one of the most concerning issues related to dark energy is in the value of the constant itself: there is, indeed, an enormous discrepancy between the estimate measured by the cosmological observations with respect to the results obtained by particle physics \cite{capozziello-delaurentis-2011} (a striking difference that can be up to 120 orders of magnitude). Also, considering the values provided by the $\Lambda$CDM model, it seems that we live in a very special epoch in which the dark energy density has the same order of magnitude as the matter density, which in general should not be the case, considering that one evolves with time while the other remains fixed: this issue is known as the coincidence problem \cite{Velten2014}. 

This problem, among others regarding also directly General Relativity (GR) from a more theoretical level, has brought the scientific community to look at alternative approaches. Among these, we find the Extended Theories of Gravity (ETG, \cite{capozziello-delaurentis-2011}), with the aim of fixing GR issues both at infrared and ultraviolet scales. For instance, the Hubble tension could be related to the GR limits at extreme scales, which in turn could be related to a certain Heisenberg principle working at cosmological scales \cite{capozziello-benetti-spallicci-2020, spallicci-benetti-capozziello-2022}.

The ETG change the framework by modifying the geometrical part of the Einstein-Hilbert action instead of the material one, with the aim of explaining the aforementioned phenomena without introducing mysterious ingredients in the cosmic density budget \cite{capozziello-faraoni-2011, blspwh11}. Indeed, the ETG have been applied both at astrophysical \cite{astashenok-capozziello-odintsov-2013, astashenok-capozziello-odintsov-2014a,astashenok-capozziello-odintsov-2014b, astashenok-capozziello-odintsov-2015a,astashenok-capozziello-odintsov-2015b, wojnar-2019,olmo-rubieragarcia-wojnar-2020,feola-etal-2020} and cosmological \cite{bahamonde-etal-2018,carloni-etal-2019,dainotti-etal-2021,saridakis-etal-2021,Benetti_2021, salucci-etal-2021} levels, with the aim of reproducing the GR results where they have been confirmed as well as extending them, overcoming the limits found at the extreme scales. Of course, these studies have direct consequences on the $\Lambda$CDM model as well.

In this paper, we instead focus on a different research line, being the presented analysis a direct following of the work performed in Refs. \cite{spallicci-etal-2021,Spallicci_2022}. This approach will be linked, at least conceptually, to the aforementioned ETG. Although we are at the dawn of the multi-messenger era thanks to the gravitational wave \cite{abbottetal2016} and neutrino \cite{gelmini-etal-2010} observations, still the majority of the information regarding our Universe has been deduced via electromagnetic detection. In our paradigm, we wonder if, as has happened in the past for GR and Newtonian gravity, a reinterpretation of light and classical electromagnetism is at hand. We indeed recall that also Special Relativity and Quantum Mechanics were born by reinterpreting light \cite{capozziello-boskoff-2021}. Thus, we shift the focus of our analysis to the messengers, the photons, rather than the astrophysical sources themselves.

Even if the classical Maxwellian theory has been proven successful, surviving the theoretical revolution of the last century, there are still some reasons why one could think of extending it, as it is shown in Ref. \cite{Spallicci_2022}. We also notice that the nature of the photon is linked to the Standard Model (SM) paradigm, according to which the photon is the only free massless particle. Thus, it is necessary to note that also the SM itself is undergoing a critical re-analysis, because of some limits which have been found by the scientific community, like the impossibility of explaining the unbalance of matter and anti-matter \cite{dibari-2022} and the neutrino masses \cite{mohapatra-senjanovic-1980,formaggio-etal-2021}. The masslessness of a particle itself cannot be proved by an experimental apparatus (no matter its accuracy), because of the Heisenberg principle in the energy-time form, which does not allow measuring any mass below $10^{-69}$ kg \cite{capozziello-benetti-spallicci-2020,spallicci-benetti-capozziello-2022}. Also, the SM fails, up to now, to provide a definite, reliable candidate for the nature of both dark energy and dark matter, as previously underlined. Because of the aforementioned reasons, the SM Extension (SME) \cite{colladay-kostelecky-1997, colladay-kostelecky-1998}, has been proposed by the scientific community. A byproduct of the SME is the possibility for the photon to acquire an effective mass proportional to the Lorentz-Poincaré Symmetry Violation (LSV) factors \cite{bonetti-dossantosfilho-helayelneto-spallicci-2017, bonetti-dossantosfilho-helayelneto-spallicci-2018}. 

Even without invoking the SME, a photon mass was firstly proposed by de Broglie in Ref. \cite{debroglie-1922}, estimated below $10^{-53}$ kg (surprisingly near the current accepted upper limits on this quantity) \cite{debroglie-1923} through dispersion analysis \cite{debroglie-1924}, for which the modified Maxwell-Amp\`ere-Faraday-Gauss equations have been written \cite{debroglie-1936,db40}. These studies were continued by his scholar Proca \cite{proca-1936b,proca-1936c,proca-1936d,proca-1937}. Regarding the experiments looking at the mass of the photon (or an upper limit related to it) in the past decades many tests have been performed, both in laboratory \cite{wifahi71} as well as considering astrophysical environments \cite{retino-spallicci-vaivads-2016, boelmasasgsp2016,boelmasasgsp2017,bebosp2017}. For a comprehensive list, see \cite{Spallicci_2022b}. For these tests, a very low radio-frequency window could be open in space in a lunar orbit away from terrestrial noise interference.

Other possible extensions of the Maxwellian theory are the Nonlinear Electromagnetism (NLEM) theories, which describe the interactions between electromagnetic fields in a vacuum. Born and Infeld solved the problem of the divergence of a point charge by relating the maximum of the electric field with the size and rest mass of the electron  \cite{born-infeld-1934a,born-infeld-1935}. Heisenberg and Euler's theory has been adopted \cite{heisenberg-euler-1936}, instead, for strong magnetic fields and in Quantum Electrodynamics. 

The important point is, whatever the non-standard electromagnetic theory considered is, a frequency shift, non-observed in the Maxwellian framework in the absence of a time or space-dependent background field, can be noted. Indeed, like the de Broglie-Proca (dBP) and SME ones, exchange energy with either galactic or intergalactic background electromagnetic fields (The SME photon exchanges also with the LSV background), thereby undergoing frequency shifts towards the red or the blue \cite{helayelneto-spallicci-2019,spallicci-etal-2021}. Also for NLEM theories, there is such an exchange and frequency shift. %\cite{helayelneto-spallicci-2022}

In \cite{Spallicci_2022}, we have investigated a frequency shift, $z_{\rm S}$, coming from these general Extended Theories of Electromagnetism (ETE) towards red or blue due to non-classical electromagnetic effects, which should be added to the cosmological expanding red shift $z_{\rm C}$. We have done so by analyzing both mock red shift as well as real data considering the Pantheon sample of Supernovae type Ia (SNe Ia) \cite{scolnic-etal-2018}, which is a catalogue of 1048 objects collected from various observational programs, as well as constraints linked to the Baryon Acoustic Oscillations (BAO) data \cite{beutler-etal-2011,blake-etal-2011,ross-etal-2015, dumasdesbourboux-etal-2020,alam-etal-2021}.

We here update this work by considering the novel Pantheon+ sample \cite{Scolnic_2022, Brout_2022}, which is a new compilation of 1701 light curves gathered from 1550 SNe Ia. This set is the natural successor of the Pantheon sample, which has been used to derive the most updated estimate on the Hubble-Lema\^itre constant, $H_0$, using the cosmological ladder approach \cite{Riess_2022}. Two additional information, one related to the uncertainties on the singular red shifts and the other on the distance moduli themselves, are present in this new catalogue, that were not presented by its precursor. Also, we note that this set does not enlarge the red shift range of its precursor, but adds more SNe Ia especially in the lower red shift regions. As we shall note, this will have a clear effect on our results. We derive new results considering this new catalogue, and compare those with the findings achieved in Ref. \cite{Spallicci_2022} while keeping the same methodology followed there, with the aim of finding possible differences in our conclusions due to the new data set considered.

In Sect. 2 we provide a synopsis of the equations introduced in 
Ref. \cite{Spallicci_2022} regarding the non-standard electromagnetism and reintroduce the methodology followed for our computations for the benefit of the reader.

In Sect. 3 we present our new results and the comparisons with those obtained in Ref. \cite{Spallicci_2022}.

In Sect. 4 we present a discussion in which we discern how it would be possible to consider both ETG and ETE for cosmological computations, thus foreseeing results taking into account both these extended frameworks. In order to do so, we present some examples in the literature where ETG have been employed to derive cosmological parameters from different data sets. Finally, the discussion and conclusions are shown in Sect. 5.

\section{Extended theories of electromagnetism  and methodology}
 In this section, we start by summarizing the discussion shown in Ref. \cite{Spallicci_2022} regarding the non-standard electromagnetic effects, in particular explaining how a further frequency shift can be deduced by these theories. Indeed, we find out that, for each of these theories, may them be NLEM, dBP, or SME, if one were to study the variation of the photon energy-momentum tensor $\theta_{\tau}^{\alpha}$ \cite{aspect-grangier-1987}, he would find terms that do not appear in the Maxwellian theory, thus leading to a more likely non-conservation of the energy tensor, which in turn would bring to a frequency shift such that
 
 \begin{equation}
\partial_\alpha \theta^\alpha_{~\tau} \longrightarrow \Delta \nu~.
 \end{equation}
The specific form of the variation depends on the theory itself \cite{spallicci-etal-2021,Spallicci_2022}, but the important point is its existence and the fact that it can be linked to a frequency variation \cite{helayelneto-spallicci-2019} for every non-standard electromagnetic effect. The frequency shifts $z_S$ are determined also by the particular hypothesis we take into account regarding their dependencies on frequencies and distance. In particular, following \cite{Spallicci_2022}, we consider four possible behaviours, $z_S$ being proportional to:

\begin{itemize}
    \item The instantaneous frequency and the distance;
    \item The emitted frequency and the distance;
    \item Only the distance;
    \item The observed frequency and the distance.
\end{itemize}

A specific $z_S$ can be associated with any cosmological measurement, being red or blue, big or small. We now show how the different shift contributions are related to one another, starting from the definition of $z = \Delta \nu/\nu_o$, where $\Delta \nu = \nu_{\rm e} - \nu_o$ is the difference between the observed $\nu_o$ and emitted 
$\nu_{\rm e}$ frequencies, or else $z = \Delta \lambda/\lambda_{\rm e}$ for the wavelengths. For the expansion, we have that $\lambda_{\rm e}$ stretches up to $\lambda_{\rm C}$ that is, $\lambda_{\rm C} = (1+z_{\rm C})\lambda_{\rm e}$. This wavelength, in turn, could be further stretched or conversely shrunk because of the ETE shift $z_{\rm S}$ to $\lambda_{\rm o} = 
(1+z_{\rm S})\lambda_{\rm C}  = (1+z_{\rm S}) (1+z_{\rm C}) \lambda_{\rm e}$. But since $\lambda_{\rm o} = (1+z)\lambda_{\rm e}$, we have 
$1 +  z = (1+z_{\rm C}) (1+z_{\rm S})$; thus
\vspace*{-0.2cm} 

\begin{eqnarray}
z = z_{\rm C} + z_{\rm S}  + z_{\rm C}z_{\rm S} \coloneqq z_{\rm o}~, 
\label{newz}
\end{eqnarray}
where $z_{\rm o}$ is the spectroscopically or photometrically observed $z$. The second-order term is non-negligible. We stress again that the expansion of the universe is represented only by the $z_{\rm C}$ red shift contribution, which will be of fundamental importance for our methodology. This red shift contribution can be derived as 

\begin{equation}
    z_{\rm C}=\frac{z-z_{\rm S}}{1+z_{\rm S}}~.
\label{zc}
\end{equation}

We also note that if $z_{\rm S}$ is negative, thus being a blue shift, the photon actually gains energy during its journey to us due to the ETE processes. This means that the expansion red shift $z_{\rm C}$ is bigger than the observed $z$, from which we deduce that the astrophysical object is actually further than what we could have detected if we were to remain in the $\Lambda$CDM model. If instead $z_{\rm S}$ is red (positive), the photon loses energy in its path because of the same ETE interactions, implying that $z_{\rm C}$ is smaller than $z$, meaning that the astrophysical object is closer to us than what it would be dictated by the same $\Lambda$CDM model. 

Regarding the relations between the ETE shift $z_S$, the distance travelled by the photon $r$ and the frequencies, we display here the related table of \cite{Spallicci_2022}, Table \ref{tabdeltanu}, in which they are all shown. 

\begin{table}
    \centering
    \begin{tabular}{|c|c|c|c|c|}
    \hline
       Type  &  1 & 2 & 3 & 4\\[6pt]\hline
         $d\nu$ & $k_{1}\nu dr$ &$k_{2}\nu_e dr$ &$k_{3} dr$ &$k_{4}\nu_o dr$ \\[6pt]\hline
         $\nu_o$ & $\nu_e e^{k_1 r}$ & $\nu_e (1+k_2 r)$ & $\nu_e + k_3 r$ & ${\ds \frac{\nu_e}{1-k_4 r}}$ \\[6pt]\hline
         $z_{\rm S}$ & $e^{-k_1 r}-1$ & ${\ds -\frac{k_2 r}{1+k_2 r}}$ & ${\ds -\frac{k_3 r}{\nu_e+k_3 r}}$ & $-k_4 r$ \\[6pt]\hline
           $k_i$ & $-{\ds \frac{\ln(1+z_{\rm S})}{r}}$  
         & $-{\ds \frac {z_{\rm S}}{r(1 + z_{\rm S})}}$ 
         & $-{\ds \frac {\nu_{\rm e} z_{\rm S}}{r(1 + z_{\rm S})}}$ 
         & $-{\ds \frac {z_{\rm S}}{r}}$ \\[6pt]\hline
         $r$ & $-{\ds \frac{\ln(1+z_{\rm S})}{k_1}}$  
         & $-{\ds \frac {z_{\rm S}}{k_2(1 + z_{\rm S})}}$ 
         & $-{\ds \frac {\nu_{\rm e} z_{\rm S}}{k_3(1 + z_{\rm S})}}$ 
         & $-{\ds \frac {z_{\rm S}}{k_4}}$ \\[6pt]\hline
         \end{tabular}
    \caption{The four different variations of the frequency $\nu$ as detailed in the text. These variations determine the observed frequency $\nu_o$, and, more importantly, the shift $z_{\rm S}$, the parameters $k_i$, and the light-travel distance $r$. }
    \label{tabdeltanu}
\end{table}

The distance $r$ chosen for these relationships is the light-travel one, being the actual length travelled by the photon in an expanding universe in its path to us, defined as

\begin{equation} \label{light travel}
    r=\frac{c}{H_0} \int_0^z \frac{dz'}{(1+z')E(z')}~,
\end{equation}
where $E(z)$, in its most general form, is 
\begin{equation} \label{E(z)}
    E(z)=\frac{H(z)}{H_0}=\sqrt{\Omega_r(1+z)^4+\Omega_M(1+z)^3+\Omega_k(1+z)^2+\Omega_{\Lambda}}~,
\end{equation}
in which we find the different density contributions in our Universe: more specifically, $\Omega_r$ is the radiation density, $\Omega_M$ the matter density (baryonic and not), $\Omega_k$ is the density associated with the curvature of the universe, and lastly $\Omega_{\Lambda}$ is the density associated with the dark energy (or cosmological constant). Another important point already investigated in Ref. \cite{Spallicci_2022} is the idea of either treating the $z_{\rm S}$ of each data point as independent one another or considering the $k_i$ parameters as general quantities, valid for every data point contemporaneously. The first assumption allows us to discern possible effects depending on features different from the distance itself, while the other allows us to treat this parameter as a cosmological one, in the same way computations involving $H_0$ or $\Omega_M$ have the aim of deriving them from a set of cosmological observations work. Both these ideas are viable, thus both have been considered in \cite{Spallicci_2022} and will be considered here.

We assume also the same cosmological models without dark energy ($\Omega_{\Lambda}=0$) introduced in Ref. \cite{Spallicci_2022}, namely:

\begin{itemize}
    \item {Cosmology model A: we set $\Omega_{M}=0.3$ and consider $\Omega_{K}=0$, implying a flat universe where the "cosmic triangle" relation $\Omega_{M}+\Omega_{K}+\Omega_{\Lambda}=1$, is not satisfied {\it a priori}. Nevertheless, the dark energy effect could be replaced {\it a posteriori} by the influence of $z_{\rm S}$. This approach supposes that, regarding the SME model, $z_{\rm S}$ is a manifestation of the LSV vacuum energy in string models, \cite{kosteleckysamuel1989b}}.
    
    \item {Cosmology model B: we take into account an open universe model, where $\Omega_{M}=0.3$ and $\Omega_{K}=0.7$, so that $\Omega_{K}+\Omega_{M}=1$ and the cosmic triangle relation is recovered.}
    
    \item {Cosmology model C: we return to the Einstein-de Sitter model \cite{einstein-desitter-1932}, which is a flat, matter-dominated universe with $\Omega_{M}=1$. This was arguably one of the most accredited cosmological models before the reintroduction of the dark energy hypothesis \cite{perlmutter-etal-1997}.}
    
\end{itemize}

For each of these models, we fix three different values of $H_0$ (in km s$^{-1}$ per Mpc) for our computations:  $H_0=67$ (consistent with the result obtained by the Planck collaboration by considering the Cosmic Microwave Background (CMB) radiation data \cite{aghanimetal2020}), $H_0=74$ (consistent with the cosmological ladder measurements \cite{riess-etal-2019}. Even if we are aware of the most updated result reached in Ref. \cite{Riess_2022} for which $H_0=73.04 \pm 1.04$, we keep $74$ as our reference value for the sake of comparisons with the findings achieved in \cite{Spallicci_2022}.) And, finally, $H_0=70$ as an average value.

We now describe the methodology behind our computations: we first note that the analysis regarding the mock red shift explored in Ref. \cite{Spallicci_2022} does not change in this work, because it is not linked to the specific data set used. Nevertheless, it can be used for comparisons as reference results.

Thus, we start our computations directly from the real data belonging to the Pantheon+ and BAO sets. For the SNe Ia data, we use the luminosity distance \cite{hogg-1999}

\begin{equation} \label{luminosity distance}
    d_{\rm L}(z)=(1+z)d_{\rm M}(z)~,
\end{equation}
where $d_{\rm M}(z)$ is the transverse comoving distance
\begin{equation} \label{comoving flat}
    d_{\rm M}(z)=\frac{c}{H_0} \int_0^z \frac{dz'}{E(z')}~,
\end{equation}
for $\Omega_{K}=0$, and
\begin{equation} \label{comoving open}
    d_{\rm M}(z)=\frac{c}{H_0 \sqrt{\Omega_{K}}} \sinh \biggl(\frac{H_0 \sqrt{\Omega_{K}}}{c} \int_0^z \frac{dz'}{E(z')} \biggr)~,
\end{equation}
for $\Omega_{K}>0$. 

For the SNe Ia data, in  Eq. (\ref{luminosity distance}) the factor $(1+z)$ refers to the heliocentric red shift, that is the apparent red shift affected by the relative motion of the Sun, and not to the red shift observed in a cosmological rest frame \cite{kenworthy-scolnic-riess-2019}. Instead, the latter appears in the integrals of Eqs. (\ref{comoving flat}, \ref{comoving open}). Both these quantities are provided by the Pantheon+ sample, and, differently from the Pantheon set, they also present their relative uncertainties. We stress again that all the cosmological distances have been computed in our approach by considering $z_C$ instead of the provided red shift $z$, which, as we shall see, has been used as a starting point for our procedure.

For both the aforementioned assumptions regarding the dependencies of the data points between one to the other, to achieve the main goal of our computations we start by using the distance-modulus, defined as (length units in Mpc)

 \begin{equation} \label{distance modulus}
        \mu=m-M=5\log_{10}\left [d_{\rm L}(z_{\rm C})\right]+25~.
\end{equation}

Differently from the Pantheon sample, where we had to fix the absolute magnitude of the SNE Ia to $-19.35$ \cite{dainotti-etal-2021, Spallicci_2022} the Pantheon+ set actually provides the values of the distance moduli themselves, obtained by using a fiducial absolute magnitude for SNE Ia derived by the SHOES 2021 Cepheids host distances \cite{Scolnic_2022, Riess_2022}. Thus, we avoid the degeneracy of the absolute magnitude with the Hubble constant, which, nevertheless, has been also investigated for this kind of analysis in \cite{Lopez-Corredoira_2022}.

For the individual computations, we minimise the following quantity for each SN Ia
\begin{equation} \label{min each SN}
    \chi_1^2=\frac{(\mu_{obs}-\mu_{th})^2}{(\mu_{obs, err})^2}~,
\end{equation}
where $\mu_{obs}$ is the distance-modulus given by the Pantheon+ sample with the corresponding diagonal error $\mu_{obs, err}$, while $\mu_{th}$ is the theoretical distance-modulus computed according to our cosmological models including  $z_{\rm S}$. As we did in \cite{Spallicci_2022}, we underline that this is not the proper $\chi^2$ of the entire Pantheon+ sample. Indeed, this computation has been performed to appreciate effects on the shift due to different SN Ia host environments, light paths, intervening electromagnetic (and LSV if applicable) fields and their alignments, which would not have been deduced by the general computation where only the distance plays the dominant role.

For the second approach, for which we consider all the SN Ia data and find a best-fit, we use the following $\chi^2$ function
\begin{equation} \label{eq_chi2_SNe}
\chi^2= (\mu_{th}-\mu_{obs})^T\times \mathcal{C}^{-1} \times (\mu_{th}-\mu_{obs})~,   
\end{equation}
where $\mathcal{C}^{-1}$ is the inverse of the covariance matrix of the Pantheon+ sample \cite{Scolnic_2022}. For our computations, we have chosen the most complete form of this covariance, accounting for all the systematic and statistical effects. For this case, a single $k_{i}$ value is computed as a best-fit parameter for all SNe Ia, which implies that $z_{\rm S}$ depends only on the light-travel distance between a specific SN Ia and us, see Table \ref{tabdeltanu}. The main advantage is the possibility of treating the $k_i$ parameters as proper, cosmological ones.

Regarding the BAO set, it is the same used in Ref. \cite{Spallicci_2022}, composed of 16 BAO-related data points. We note that behind each of these points there are hundred of thousand of observations of galaxy correlations, Lyman $\alpha$ forests, and quasars \cite{alam-etal-2021}. As in Ref. \cite{Spallicci_2022}, because of their nature, we have decided to include them only in the general fit computations, and not in the individual derivations, in which only SNe Ia are involved.

We also have to stress that, while the SNe Ia observations are all associated with the distance-modulus and, thus, with the luminosity distance, this is not the case for the BAO. Indeed, the quantities measured using them are related through the following equations

\begin{equation}
        d_V(z)=\biggr[{d_M}^2(z)\frac{cz}{H(z)} \biggr]^\frac{1}{3},
    \label{eq_dilationscale}
    \end{equation}
    \begin{equation}
        A(z)=\frac{100 d_V(z)\sqrt{\Omega_M h^2}}{cz},
        \label{Aparameter}
    \end{equation}
    \begin{equation}
        d_H(z)=\frac{c}{H(z)},
        \label{dH}
    \end{equation}
    and the comoving distance defined in Eqs. (\ref{comoving flat}, \ref{comoving open}). Here, $h=H_0/100$ km s$^{-1}$ per Mpc. We also stress that in Ref. \cite{alam-etal-2021} the results reported concerning these detected quantities are rescaled by the sound horizon $r_d$, for which, in our computations, we have used the following numerical approximation \cite{aubourg-etal-2015,sharov-2016}
    
    \begin{equation}
    r_d=\frac{55.154 \cdot e^{[-72.3(\Omega_{\nu}h^{2}+0.0006)^2]}}{(\Omega_{M}h^{2})^{0.25351}(\Omega_{b}h^{2})^{0.12807}}Mpc~,
    \label{eq_rsfiducialtrue}
    \end{equation}
    where $\Omega_{b}$ is the baryonic density in the universe, set to the value $\Omega_{b}\cdot h^2=0.02237$ \cite{aghanimetal2020}, while $\Omega_{\nu}$ is the neutrino density in the universe, fixed to the value provided by the $\Lambda$CDM model $\Omega_{\nu} \cdot h^2=0.00064$.
    
    Regarding the $\chi^2$ associated with the BAO, we considered each subset of the total BAO sample, given the different quantities measured and the possible covariance terms between them \cite{beutler-etal-2011, blake-etal-2011,ross-etal-2015,dumasdesbourboux-etal-2020, alam-etal-2021}. The forms of these functions are the same as Eq. (\ref{eq_chi2_SNe}).

As in Ref. \cite{Spallicci_2022}, the computations themselves have been carried out by taking into account a Bayesian approach, using the Cobaya tool \cite{torrado-lewis-2021}, a code for Bayesian analysis in Python, that adopts a Markow-Chain Monte-Carlo (MCMC) method.

\section{Results}
We now show the results obtained by our new computations. As previously mentioned, the mock red shift derivations are not affected by the new Pantheon+ set, thus no new results have been achieved using them, and we refer directly to the calculations performed in Ref. \cite{Spallicci_2022} for our comparisons.

\subsection{Individual SNe Ia results}

\begin{figure}
    \centering
    \includegraphics[width=0.33\hsize,height=0.3\textwidth,angle=0,clip]{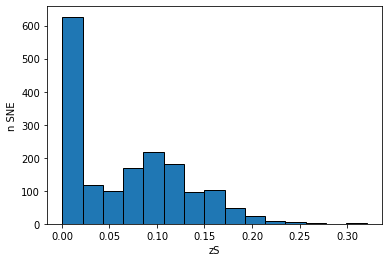}
    \includegraphics[width=0.33\hsize,height=0.3\textwidth,angle=0,clip]{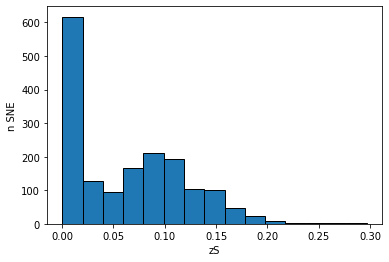}
    \includegraphics[width=0.33\hsize,height=0.3\textwidth,angle=0,clip]{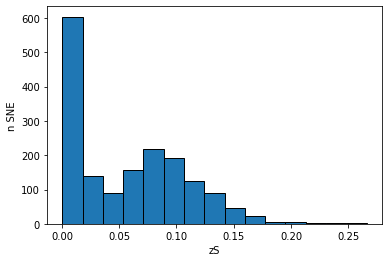}
    \includegraphics[width=0.33\hsize,height=0.3\textwidth,angle=0,clip]{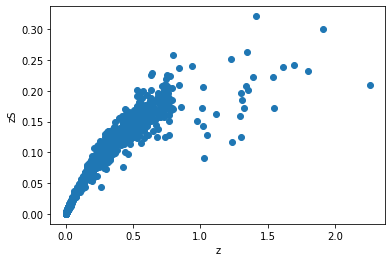}
    \includegraphics[width=0.33\hsize,height=0.3\textwidth,angle=0,clip]{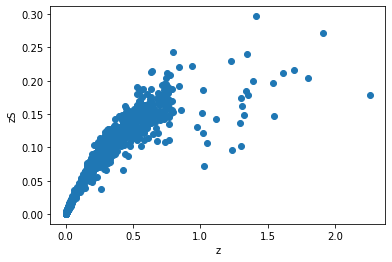}
    \includegraphics[width=0.33\hsize,height=0.3\textwidth,angle=0,clip]{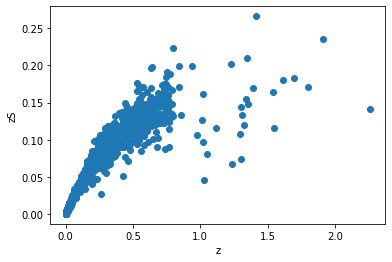}
    \caption{The first row shows the histograms of $z_{\rm S}$ for the Cosmology model A, where $\Omega_{M}=0.3$, $\Omega_{k}=\Omega_{\Lambda}=0$, related to the Pantheon+ sample. The second row shows instead the scatter plot $z_{\rm S}$ versus $z$. We fixed the values for $H_0$ to 67 (first column), 70 (second column), 74 (third column), km s$^{-1}$ per Mpc.}
    \label{fig:firstthreecasessingular}
\end{figure}

\begin{figure}
    \centering
    \includegraphics[width=0.33\hsize,height=0.3\textwidth,angle=0,clip]{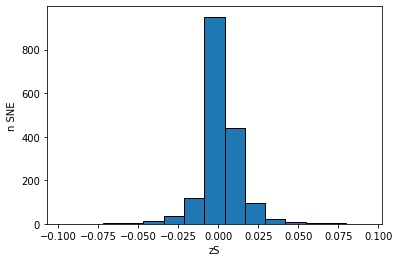}
    \includegraphics[width=0.33\hsize,height=0.3\textwidth,angle=0,clip]{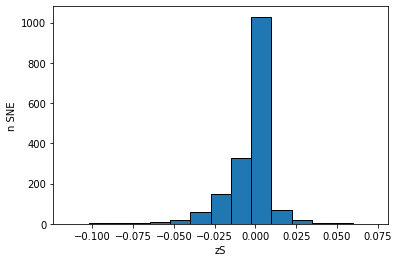}
    \includegraphics[width=0.33\hsize,height=0.3\textwidth,angle=0,clip]{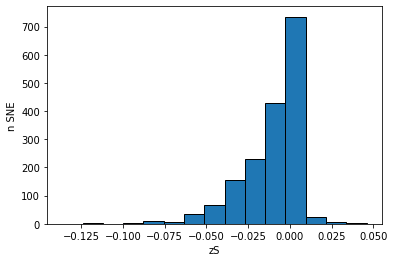}
    \includegraphics[width=0.33\hsize,height=0.3\textwidth,angle=0,clip]{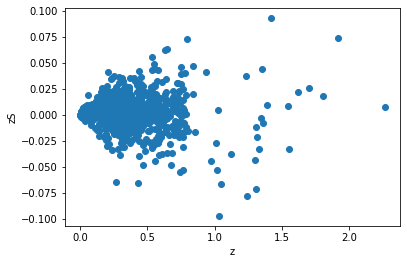}
    \includegraphics[width=0.33\hsize,height=0.3\textwidth,angle=0,clip]{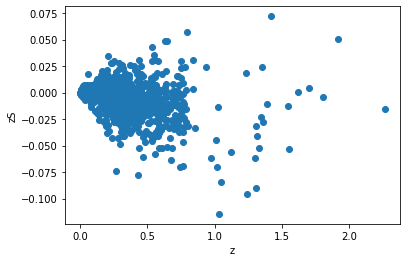}
    \includegraphics[width=0.33\hsize,height=0.3\textwidth,angle=0,clip]{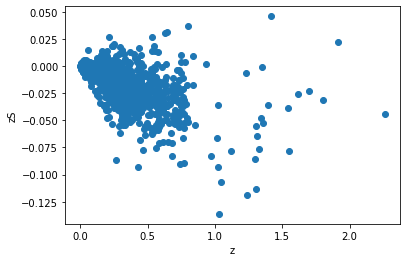}
    \caption{The first row shows the histograms of $z_{\rm S}$ for the Cosmology model B, where $\Omega_{M}=0.3$, $\Omega_{k}=0.7$, and $\Omega_{\Lambda}=0$, related to the Pantheon+ sample. The second row shows the scatter plot $z_{\rm S}$ versus $z$.  We fixed the values for $H_0$ to 67 (first column), 70 (second column), 74 (third column), km s$^{-1}$ per Mpc.}
    \label{fig:secondthreecasessingular}
\end{figure}

\begin{figure}
    \centering
    \includegraphics[width=0.33\hsize,height=0.3\textwidth,angle=0,clip]{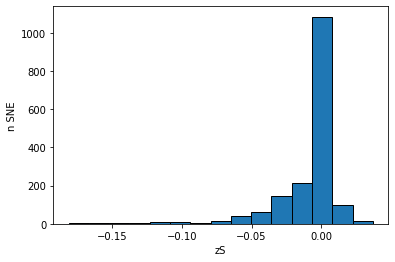}
    \includegraphics[width=0.33\hsize,height=0.3\textwidth,angle=0,clip]{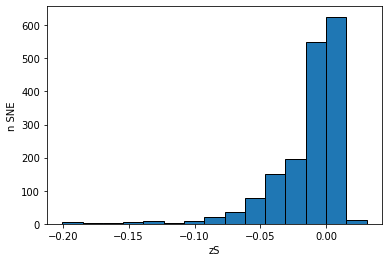}
    \includegraphics[width=0.33\hsize,height=0.3\textwidth,angle=0,clip]{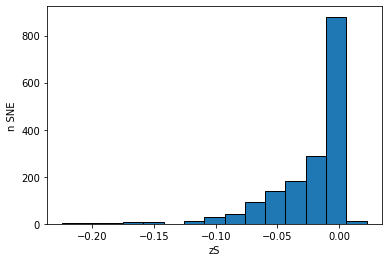}
    \includegraphics[width=0.33\hsize,height=0.3\textwidth,angle=0,clip]{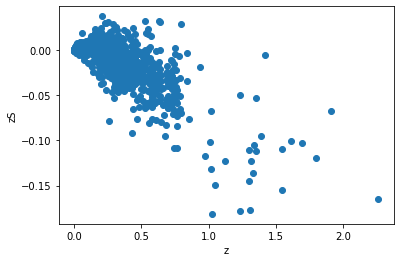}
    \includegraphics[width=0.33\hsize,height=0.3\textwidth,angle=0,clip]{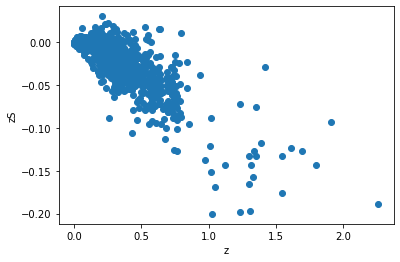}
    \includegraphics[width=0.33\hsize,height=0.3\textwidth,angle=0,clip]{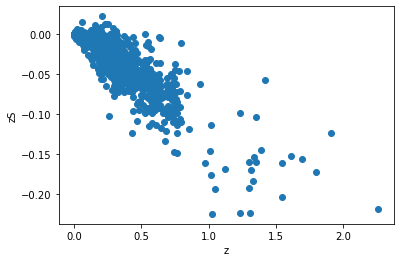}
    \caption{The first row shows the histograms of $z_{\rm S}$ for the Cosmology model C, where $\Omega_{M}=1$, $\Omega_{k}=\Omega_{\Lambda}=0$, related to the Pantheon+ sample. The second row shows the scatter plot $z_{\rm S}$ versus $z$. We fixed the values for $H_0$ to 67 (first column), 70 (second column), 74 (third column), 
    km s$^{-1}$ per Mpc.}
    \label{fig:thirdthreecasessingular}
\end{figure}

We here show the results obtained for the individual SNe Ia computations considering the Pantheon+ sample, which are gathered in Figs. \ref{fig:firstthreecasessingular}, \ref{fig:secondthreecasessingular}, and \ref{fig:thirdthreecasessingular}. In each of these figures, we show the histograms of $z_{\rm S}$ for the three Cosmology models and the plots of $z_{\rm S}$ versus $z$. 

Let us start from Fig. \ref{fig:firstthreecasessingular}, where the results for the Cosmology model A, in which we recall $\Omega_{M}=0.3$, $\Omega_{k}=\Omega_{\Lambda}=0$, are visualized. In the first row, where the histograms of the obtained $z_{\rm S}$ are shown, we note a peak for $z_{\rm S}$ consistent with zero, followed by a distribution presenting a secondary peak, which is consistent or slightly below $z_{\rm S}=0.1$, depending on the value for $H_0$. The maximum value reached for $z_{\rm S}$ also depends on $H_0$. We note that in these computations $z_{\rm S}$ is always positive for every value of $H_0$.

Comparing these histograms with the results obtained in Ref. \cite{Spallicci_2022} for the same case, we highlight the following features:
\begin{itemize}
    \item In both cases, $z_{\rm S}$ is always positive, confirming also the results obtained with the mock red shift for the same cosmology model.
    \item In \cite{Spallicci_2022}, the peak of $z_{\rm S}$ consistent with zero was also present, but was not as predominant as in the new results, being of the same magnitude as the secondary peak.
    \item The new distribution is wider: both the secondary peak as well as the maximum value of $z_{\rm S}$ are bigger than the same quantities derived by the Pantheon set, and this is true for all the values of $H_0$.
    \item the dependency on the value of $H_0$ is very similar between the two results: increasing $H_0$ decreases both the value of the secondary peak as well as of the maximum.
\end{itemize}

In the second row, instead, the scatter plot visualizing the behaviour of $z_{\rm S}$ with respect to the detected $z$ is shown. We note a dispersion probably due to the errors in the distance modulus, which is similar to the one noted in Ref. \cite{Spallicci_2022}. Indeed, we notice that in general the scatter plots obtained with the Pantheon sample are similar to the new computations. We thus conclude that the most relevant difference between the two results is the peak of low $z_{\rm S}$, which is dominant in the new histograms obtained using the Pantheon+ set. A possible reason would be the nature of the new data used in the Pantheon+ sample, which are, in general, found at low detected red shifts.

We now move to Fig. \ref{fig:secondthreecasessingular}, in which we take into account the Cosmology model B, where $\Omega_{M}=0.3$, $\Omega_{k}=0.7$, and $\Omega_{\Lambda}=0$. The main feature of the histograms shown in the first row is the predominant peak for $z_{\rm S}$ around 0. We also note a very interesting behaviour depending on the fixed value of $H_0$: indeed, for $H_0=67$  km s$^{-1}$ per Mpc, the distribution is fairly symmetric around 0, while increasing the value of $H_0$ breaks this symmetry, having a bigger tail of negative values for $z_{\rm S}$. The result for $H_0=67$  km s$^{-1}$ per Mpc is particularly interesting: indeed, a $z_{\rm S}$ consistent with zero would imply that the open cosmological model without dark energy is able to reproduce the results of the $\Lambda$CDM model without invoking the $z_{\rm S}$ contribution, even if we see also a large dispersion of the computed values, from which we deduce that the effect due to $z_{\rm S}$ is not negligible.

Comparing these results with the computations performed for the corresponding case in Ref. \cite{Spallicci_2022}, we note that we did not find the same symmetric behaviour with the Pantheon sample: indeed, the majority $z_{\rm S}$ was negative even for $H_0=67$  km s$^{-1}$ per Mpc. We also note that the intervals of the $z_{\rm S}$ obtained in the two computations have comparable widths, with the Pantheon+ results being more shifted to positive values with respect to the corresponding Pantheon findings. We also note the peak near $z_{\rm S}=0$, to which we give the same explanation concerning the similar finding observed for Fig. \ref{fig:firstthreecasessingular}.

On the second row we see the scatter plots, similarly to what we have shown in Fig. \ref{fig:firstthreecasessingular}. It is clear from this plot the symmetric behaviour for $H_0=67$  km s$^{-1}$, that is broken by increasing the value of $H_0$. We also note that the negative values of $z_{\rm S}$ seem to increase with the red shift, especially for $H_0=70$  km s$^{-1}$ per Mpc and $H_0=74$  km s$^{-1}$ per Mpc. Finally, we also note that the scattering characterizing these plots seems to decrease with $H_0$.

We now discuss the results obtained in Fig. \ref{fig:thirdthreecasessingular} for the Cosmology model C, where we fixed $\Omega_{M}=1$, $\Omega_{k}=\Omega_{\Lambda}=0$. Looking at the first row, where the histograms of $z_{\rm S}$ are shown, we note that the majority of the $z_{\rm S}$ seem to be negative, as was the case for the same computations for the Pantheon sample. In reality, as we will note later, when we compute the $k_i$ parameters, the number of negative and positive $z_{\rm S}$ is closer than what we would expect from these panels, especially for $H_0=67$  km s$^{-1}$ per Mpc. This trend is concordant with the value of $H_0$, as again was the case for the previous computations. Indeed, in general, we can see that this is the particular comparison in which the results obtained for the Pantheon+ and Pantheon sets are more alike. The main difference is in the peak near $z_{\rm S}=0$, as in the other computations, to which we give the same origin as in the previous two cases. In the second row, we note from the scatter plots that $z_{\rm S}$ becomes more negative at high red shifts, as we obtained for the computations performed in Ref. \cite{Spallicci_2022}. Indeed, even from this plot, we can appreciate the similarity between the two results (and also with the mock red shift for the same cosmological model). Regarding the maximum (in absolute value) reached by $z_{\rm S}$, we note that it increases with the value of $H_0$, and also that it is bigger for the Pantheon sample than the Pantheon+ value.

Given that the new Pantheon+ sample provides also an estimate of the errors on the observed red shifts, we can compare our results on $z_{\rm S}$ with them, in order to understand if, by considering the uncertainties on the measurements, a similar effect to the one deduced by us for the non-standard shift can be achieved by the observational errors themselves. We thus compute the ratio between $z_{\rm S}$ and the errors, from where we derive the median of the distributions. The results are gathered in Table \ref{tab:medianSingular}, in which we note that the $z_{\rm S}$ values are consistently predominant over the uncertainties provided by the Pantheon+ sample, which implies that their effects cannot be accounted for by errors in the measurements.

 \begin{table}
    \centering
    \begin{tabular}{c|c|c|c}
    \hline
       Cosmology  &  $H_0=67$& $H_0=70$ & $H_0=74$ \\\hline
         Cosmology A & $32.1$ & $30.14$ &$27.8$ \\\hline
         Cosmology B & $3.06$ &$2.51$ &$3.39$  \\\hline
         Cosmology C, & $3.25$ &$3.38$ & $5.37$ \\\hline
    \end{tabular}
    \caption{The median of the distributions of the ratios between the computed $z_{\rm S}$ and the uncertainties provided by the Pantheon+ sample. The values for $H_0$ are in km s$^{-1}$ per Mpc.} 
    \label{tab:medianSingular}
\end{table}

As we did in Ref. \cite{Spallicci_2022}, from these computations and using the relations in Table \ref{tabdeltanu}, we can now compute the $k_i$ parameters for each SN Ia belonging to the Pantheon+ sample. This has been done for every cosmological model and value for $H_0$ previously considered. The results are illustrated in Table \ref{tab:resultskiSingular}, where we show the mean and the standard deviation of the $k_i$ distributions. For our computations we choose for the emitted frequency $k_{3}$, $\nu_{e}=6.74 \times 10^{14} s^{-1}$, which is an optical frequency in the B-Band. 

  \begin{table}
    \centering
    \begin{tabular}{c|c|c|c|c}
    \hline
       Cosmology  &  $k_1$ & $k_2$ & $k_3$ & $k_4$\\\hline
         Cosmology A, $H_0=67$ & $(-9.94 \pm 2.46) \times 10^{-5}$ &$(-9.67 \pm 2.56) \times 10^{-5}$ &$(-6.48 \pm 1.71) \times 10^{10}$ &$(-1.02 \pm 0.24) \times 10^{-4}$ \\\hline
         Cosmology A, $H_0=70$ & $(-9.45 \pm 2.44 )\times 10^{-5}$ &$(-9.20 \pm 2.52 )\times 10^{-5}$ &$(-6.16 \pm 1.69)\times 10^{10}$ &$(-9.71 \pm 2.36 )\times 10^{-5}$ \\\hline
         Cosmology A, $H_0=74$ & $(-8.77 \pm 2.41 )\times 10^{-5}$ &$(-8.57 \pm 2.47) \times 10^{-5}$ &$(-5.74 \pm 1.66) \times 10^{10}$ &$(-9.00 \pm 2.35 )\times 10^{-5}$ \\\hline
         Cosmology B, $H_0=67$ & $(-1.01 \pm 1.95 )\times 10^{-5}$ &$(-1.00 \pm 1.95)\times 10^{-5}$ &$(-0.67 \pm 1.31) \times 10^{10}$ &$(-1.02 \pm 1.95 )\times 10^{-5}$ \\\hline
         Cosmology B, $H_0=70$ & $(-0.06 \pm 1.95 )\times 10^{-5}$ &$(-0.05 \pm 1.95)\times 10^{-5}$ &($-0.03 \pm 1.31) \times 10^{10}$ &($-0.06 \pm 1.94) \times 10^{-5}$ \\\hline
         Cosmology B, $H_0=74$ & $(1.22 \pm 1.94) \times 10^{-5}$ &$(1.24 \pm 1.96)\times 10^{-5}$ &$(0.83 \pm 1.31) \times 10^{10}$ &($1.21 \pm 1.93 )\times 10^{-5}$ \\\hline
         Cosmology C, $H_0=67$ & $(-0.19 \pm 2.43 )\times 10^{-5}$ &$(-0.17 \pm 2.46 )\times 10^{-5}$ &$(-0.12 \pm 1.65) \times 10^{10}$ &($0.20 \pm 2.40 )\times 10^{-5}$ \\\hline
         Cosmology C, $H_0=70$ & $(0.84 \pm 2.46 )\times 10^{-5}$ &$(0.87 \pm 2.52)\times 10^{-5}$ &($0.59 \pm 1.69) \times 10^{10}$ &($0.81 \pm 2.42) \times 10^{-5}$ \\\hline
          Cosmology C, $H_0=74$ & ($2.23 \pm 2.52) \times 10^{-5}$ &($2.29 \pm 2.62)\times 10^{-5}$ &($1.53 \pm 1.75) \times 10^{10}$ &($2.17 \pm 2.44 )\times 10^{-5}$ \\\hline
    \end{tabular}
    \caption{Mean and standard deviation values of the $k_i$ parameters derived from the individual computations of $z_{\rm S}$, for the three cosmological models. The computations have been performed considering the distances in Mpc, which means that the $k_i$ parameters are in Mpc$^{-1}$ for $i=1,2,4$ and Mpc$^{-1}$ s$^{-1}$ for $k=3$, while the values for $H_0$ are in km s$^{-1}$ per Mpc, as usual.} 
    \label{tab:resultskiSingular}
\end{table}

The derived values for the $k_i$ parameters present the same order of magnitude with respect to the results found in Ref. \cite{Spallicci_2022}: indeed, we note that all the new findings are consistent with the derivations achieved with the Pantheon set within 1 $\sigma$. As in the old results, the computed standard deviation is of the same order of magnitude (or even higher) than the mean values, and we obtain results consistent with zero for the cosmology models B and C. This was expected from the plots shown beforehand, because of the presence of both positive and negative $z_{\rm S}$ and because of the general dispersion observed in the real data.

\subsection{General best-fit}
We now move to the second approach, in which we have computed the $k_i$ considering a general best-fit procedure. We first consider only the SNe Ia data, to which we later add the BAO. To compute these parameters, we follow the same iterative algorithm used in Ref. \cite{Spallicci_2022}, namely, we use Eqs. (\ref{zc}, \ref{light travel} , \ref{luminosity distance}) and Table \ref{tabdeltanu} in the following way

\[ z\rightarrow r (z)\rightarrow k_i (z) \rightarrow z_{\rm S} (z) \rightarrow \\
z_{\rm C} (z) \rightarrow r (z_{\rm C}) \rightarrow k_i (z_{\rm C}) \rightarrow {\rm and \: back \: again}~.
\]
More precisely, we start from the observed red shift $z$ shown in the catalogue, from which we compute the light travel distance, that in turn, using our Bayesian best-fit derivation, has been used to calculate the best-fit values for the $k_i$ parameters via Eq. (\ref{eq_chi2_SNe}). These parameters have then been considered to compute $z_{\rm S}$ and $z_{\rm C}$. Once we derive $z_{\rm C}$, we go back to the light travel distance and the best-fit computation for the $k_i$ using this new cosmological red shift instead of the observed one, and we repeat the procedure. We stop the computations once the difference between the light-travel distances computed in two subsequent steps becomes small enough (in particular, a mean difference between these two values of $\sim 1 Mpc$ considering all SNe Ia). The same has been done for the BAO. 

As for the previous computations, we gather 9 cases (3 Cosmology models, 3 fixed values of $H_0$), from which we derive our estimations for the $k_i$ best-fits. Regarding the Bayesian priors, as in Ref. \cite{Spallicci_2022} we have chosen flat priors, with interval ranges varying case by case, taking as a baseline the results computed in the previous section regarding the individual SNe Ia. The derivations obtained for the $k_{i}$ parameters are shown in Table \ref{tab:resultskigeneral}. 

\begin{table}
    \centering
    \begin{tabular}{c|c|c|c|c}
    \hline
       Cosmology  &  $k_1$ & $k_2$ & $k_3$ & $k_4$\\\hline
         Cosmology A, $H_0=67$ & $(-9.84\pm 0.69)  \times 10^{-5}$ &$(-8.85 \pm 0.69) \times 10^{-5}$ &$(-5.88\pm 0.41 ) \times 10^{10}$ &$(-1.05\pm 0.08)  \times 10^{-4}$ \\\hline
         Cosmology A, $H_0=70$ & $(-9.38\pm 0.73)  \times 10^{-5}$ &$(-8.47 \pm 0.63) \times 10^{-5}$ &$(-5.66\pm 0.45 ) \times 10^{10}$ &$(-1.01\pm 0.09)  \times 10^{-4}$ \\\hline
         Cosmology A, $H_0=74$ & $(-8.70\pm 0.73)  \times 10^{-5}$ &$(-7.93 \pm 0.74) \times 10^{-5}$ &$(-5.27\pm 0.46 ) \times 10^{10}$ &$(-9.17\pm 0.86)  \times 10^{-5}$ \\\hline
        Cosmology B, $H_0=67$ & $(-0.68\pm 1.21)  \times 10^{-5}$ &$(-0.67 \pm 0.97) \times 10^{-5}$ &$(-3.63\pm 7.46 ) \times 10^{9}$ &$(-0.70\pm 1.22)  \times 10^{-5}$ \\\hline
         Cosmology B, $H_0=70$ & $(0.26\pm 1.22)  \times 10^{-5}$ &$(0.26 \pm 1.21) \times 10^{-5}$ &$(2.20\pm 8.69 ) \times 10^{9}$ &$(0.27\pm 1.33)  \times 10^{-5}$ \\\hline
         Cosmology B, $H_0=74$ & $(1.54\pm 1.27)  \times 10^{-5}$ &$(1.53 \pm 1.35) \times 10^{-5}$ &$(1.06\pm 0.87 ) \times 10^{10}$ &$(1.55\pm 1.24)  \times 10^{-5}$ \\\hline
         Cosmology C, $H_0=67$ & $(-0.24\pm 1.13)  \times 10^{-5}$ &$(-0.23 \pm 1.31) \times 10^{-5}$ &$(0.07\pm 7.64 ) \times 10^{9}$ &$(0.56\pm 1.22)  \times 10^{-5}$ \\\hline
        Cosmology C, $H_0=70$ & $(0.77\pm 1.43)  \times 10^{-5}$ &$(0.74 \pm 1.20) \times 10^{-5}$ &$(5.57\pm 7.91 ) \times 10^{9}$ &$(0.80\pm 1.29)  \times 10^{-5}$ \\\hline
          Cosmology C, $H_0=74$ & $(2.12\pm 1.59)  \times 10^{-5}$ &$(2.05 \pm 1.69) \times 10^{-5}$ &$(1.37\pm 1.02 ) \times 10^{10}$ &$(2.19\pm 1.61)  \times 10^{-5}$ \\\hline
    \end{tabular}
    \caption{
   The best-fit values with the relative errors for the $k_i$ parameters, considering the general best-fit for all the SNe Ia belonging to the Pantheon+ sample, for the three cosmological models. The computations have been performed considering the distances in Mpc, which means that the $k_i$ parameters are in Mpc$^{-1}$ for $i=1,2,4$ and Mpc$^{-1}$ s$^{-1}$ for $k=3$, while the values for $H_0$ are in km s$^{-1}$ per Mpc.} 
    \label{tab:resultskigeneral}
\end{table}

We first note how the results, albeit having different meanings, shown in Tables \ref{tab:resultskiSingular} and \ref{tab:resultskigeneral} are consistent within 1 $\sigma$, as was the case for the Pantheon computations. We also note how the uncertainties for the general case are, in general, significantly smaller than the scatters derived for the computations in which we have considered the singular SNe Ia, still reminding the conceptual difference between the two quantities. 

We now compare this table with the corresponding results derived in Ref. \cite{Spallicci_2022}. We note how, contrary to what we obtained for Table \ref{tab:resultskiSingular}, not all the new $k_i$ parameters derived in this work are consistent with the old findings: indeed, a direct comparison expressing in how many $\sigma$ the values are (identified by the sum of the errors on the two results) is shown in Table \ref{tab:Comaprisons}. We note that the highest discrepancies have been found for the Cosmology model A, while in the Cosmology model B the results are almost always consistent within 1 $\sigma$. An interesting difference regards the results with $H0=67$ km s$^{-1}$ per Mpc for the cosmology models B and C: indeed, we note how the old results presented a positive mean value, while the new results display a negative mean. This would imply a change of the sign for the related $z_{\rm S}$, as we will see. This difference could be due to the following dichotomy, namely the fact that there are more data points consistent with a negative value for $z_{\rm S}$ in the Pantheon+ sample, as well as because the new results for the best-fit values are consistent with zero.

Indeed, another key difference between the old and new derivations is in the magnitude of the related uncertainty: for the new results this is significantly higher than what we obtained in Ref. \cite{Spallicci_2022}. A possible reason for this conclusion is the higher number of probes in the new set, which would bring to a higher possible variability for the $z_{\rm S}$, even if such uncertainty is still considerably smaller than the scatter derived for the individual SN Ia case (keeping in mind the different meanings between the two quantities).

\begin{table}
    \centering
    \begin{tabular}{c|c|c|c|c}
    \hline
       Cosmology  &  $k_1$ & $k_2$ & $k_3$ & $k_4$\\\hline
         Cosmology A, $H_0=67$ & $2.32$ &$1.56$ &$1.69$ &$2.37$ \\\hline
         Cosmology A, $H_0=70$ & $2.26$ &$1.8$ &$1.64$ &$2.26$ \\\hline
         Cosmology A, $H_0=74$ & $2.22$ &$1.59$ &$1.62$ &$2.14$ \\\hline
        Cosmology B, $H_0=67$ & Below $1$ &$1.05$ & Below $1$ & Below $1$ \\\hline
         Cosmology B, $H_0=70$ & Below $1$ & Below $1$ & Below $1$ & Below $1$ \\\hline
         Cosmology B, $H_0=74$ &  Below $1$ &  Below $1$ & Below $1$ & Below $1$\\\hline
         Cosmology C,$H_0=67$ & $1.50$ &$1.31$ &$1.29$ & Below $1$ \\\hline
        Cosmology C, $H_0=70$ & $1.22$ &$1.48$ &$1.43$ &$1.31$ \\\hline
          Cosmology C, $H_0=74$ & $1.13$ &$1.13$ &$1.25$ &$1.03$ \\\hline
    \end{tabular}
    \caption{
   Table of the comparisons between the results shown in Table \ref{tab:Comaprisons} and the corresponding findings obtained in Ref. \cite{Spallicci_2022}. The numbers shown are the ratio between the difference between the two mean value and the sum of the corresponding errors. If this number is below $1$ (i. e., the two results are consistent) we just show "Below $1$" in the table.} 
    \label{tab:Comaprisons}
\end{table}

\begin{figure}
    \centering
    \includegraphics[width=0.33\hsize,height=0.3\textwidth,angle=0,clip]{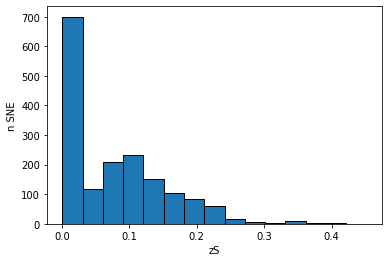}
    \includegraphics[width=0.33\hsize,height=0.3\textwidth,angle=0,clip]{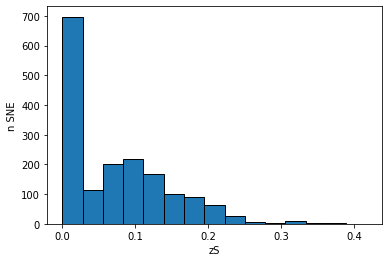}
    \includegraphics[width=0.33\hsize,height=0.3\textwidth,angle=0,clip]{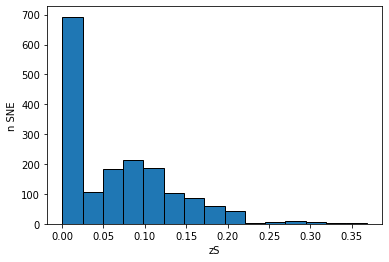}
    \caption{Histograms of the computed $z_{\rm S}$ from the $k_1$ parameter considering the Cosmology model A, where  $\Omega_{M}=0.3$, $\Omega_{k}=\Omega_{\Lambda}=0$ ($H_0=67, 70, 74$ for the left, central and right panels. The values for $H_0$ are in km s$^{-1}$ per Mpc).}
    \label{fig:firstthreecasesgeneralredshift}
\end{figure}

\begin{figure}
    \centering
    \includegraphics[width=0.33\hsize,height=0.3\textwidth,angle=0,clip]{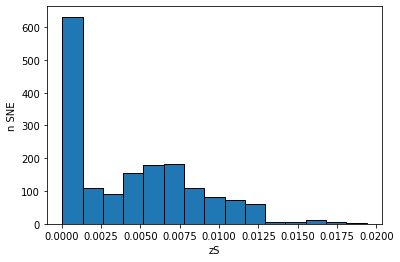}
    \includegraphics[width=0.33\hsize,height=0.3\textwidth,angle=0,clip]{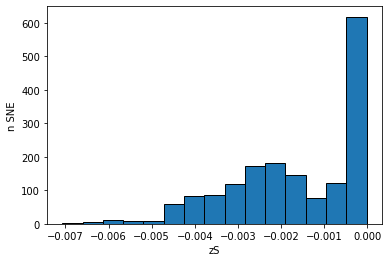}
    \includegraphics[width=0.33\hsize,height=0.3\textwidth,angle=0,clip]{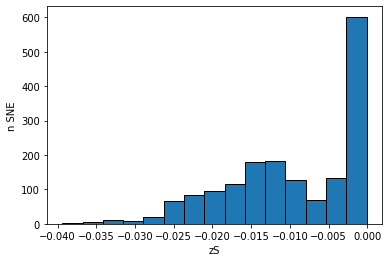}
    \caption{Histograms of the computed $z_{\rm S}$ from the $k_1$ parameter considering the Cosmology model B, where $\Omega_{M}=0.3$, $\Omega_{k}=0.7$, $\Omega_{\Lambda}=0$ ($H_0=67, 70, 74$ for the left, central and right panels. The values for $H_0$ are in km s$^{-1}$ per Mpc).}
    \label{fig:secondthreecasesgeneralredshift}
\end{figure}

\begin{figure}
    \centering
    \includegraphics[width=0.33\hsize,height=0.3\textwidth,angle=0,clip]{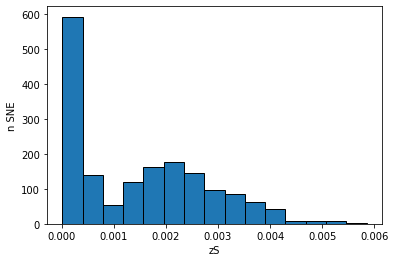}
    \includegraphics[width=0.33\hsize,height=0.3\textwidth,angle=0,clip]{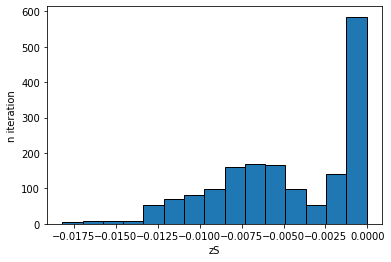}
    \includegraphics[width=0.33\hsize,height=0.3\textwidth,angle=0,clip]{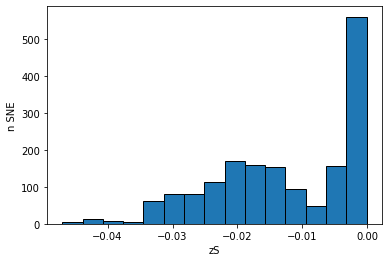}
    \caption{
    Histograms of the computed $z_{\rm S}$ from the $k_1$ parameter considering the Cosmology model C, where $\Omega_{M}=1$, $\Omega_{k}=\Omega_{\Lambda}=0$ ($H_0=67, 70, 74$ for the left, central and right panels. The values for $H_0$ are in km s$^{-1}$ per Mpc).}
    \label{fig:thirdthreecasesgeneralredshift}
\end{figure}

From the $k_i$ parameters we can now derive the $z_{\rm S}$ for the entire data set, as we did in Ref. \cite{Spallicci_2022}. The results are shown in Figs. \ref{fig:firstthreecasesgeneralredshift}, \ref{fig:secondthreecasesgeneralredshift}, and \ref{fig:thirdthreecasesgeneralredshift}, where the histograms for the $z_{\rm S}$ in the general case are shown. Again, we show the results related to $k_1$, because no significant difference has been observed for the $k_i$ parameters even with the new set.

Let us comment on these histograms starting from Fig. \ref{fig:firstthreecasesgeneralredshift}, where the results related to the Cosmology model A are visualized. As we expected from the formula inside Table \ref{tabdeltanu}, the $z_{\rm S}$ derived from this model are always positive for every value of $H_0$. We again note the peak for $z_{\rm S} \sim 0$, which is due to the additional low red shift SNe Ia present in the Pantheon+ sample. We also note that increasing $H_0$ decreases the maximum value obtained for $z_{\rm S}$, as was the case for the individual computations, of which they keep the same general order of magnitude. This feature was present also in the results obtained in \cite{Spallicci_2022}.

We now move to the histograms visualized in Fig. \ref{fig:secondthreecasesgeneralredshift}, which represent the results related to the Cosmology model B. We note that the $z_{\rm S}$ are rather small, and that for $H_0=67$ km s$^{-1}$ per Mpc they are positive, while for the other two values they are negative. This depends on the mean value of the general parameter $k_1$, which was negative for $H_0=67$ and positive for $H_0=70, 74$ km s$^{-1}$ per Mpc. This effect has been relatively hinted from the corresponding individual results, for which a rather symmetric distribution around $z_{\rm S}=0$ has been found for $H_0=67$ km s$^{-1}$ per Mpc, that has then moved to more negative values for $H_0=70, 74$ km s$^{-1}$ per Mpc. Given how the $z_{\rm S}$ values are computed in this second approach, we can only have either an entirely positive or negative set of results for $z_{\rm S}$ for a singular histogram. This feature was not observed for the results derived from the older Pantheon sample, for which negative values for $z_{\rm S}$ were observed also for $H_0=67$ km s$^{-1}$ per Mpc. We also note that the $z_{\rm S}$ for the new computations are, in general, smaller than the values obtained in the Pantheon derivations. Regarding the changing sign, this could be because of the symmetric behaviour observed in the new results (which is shown also from the general low values for $z_{\rm S}$), not observed with the old Pantheon sample, that is broken by increasing the value of $H_0$. We finally notice the usual peak at $z_{\rm S}=0$ due to the new low red shift SNe Ia.

We now comment on the results shown in Fig. \ref{fig:thirdthreecasesgeneralredshift}, where we find the Cosmology model C. We may note the similarities with respect to the results obtained for the Cosmology model B, albeit here we also notice how the $z_{\rm S}$ are usually larger. Indeed, we see that for $H_0=67$ km s$^{-1}$ per Mpc we obtain positive values for $z_{\rm S}$, while we find negative parameters for $H_0=70,74$ km s$^{-1}$ per Mpc. Again, this difference has not been noted either for the individual SNe Ia results nor for the derivations in Ref. \cite{Spallicci_2022}. The similarity of this difference with the results achieved for the Cosmology model B allows us to conclude that the same reasoning deduced previously applies also here. In general, we note that the relation between the computed $z_{\rm S}$ and the fixed value of $H_0$ is the same as the dependence observed for the old Pantheon sample for all the Cosmology models considered.

As we did for the individual computations, we can now compare our $z_{\rm S}$ with the correspondent uncertainties provided by the Pantheon+ sample. The ratios for the $k_1$ parameter are shown in Table \ref{tab:medianGlobal}. We observe again that, in general, the computed values of $z_{\rm S}$ are higher than the observational errors, with the only exceptions for Cosmology B, $H_0=70$ km s$^{-1}$ per Mpc, and  Cosmology C, $H_0=67$ km s$^{-1}$ per Mpc, where the median displays that the two quantities are of the same order of magnitude. This is due to the small values for the $k_1$ parameters found in those specific cases.

\begin{table}
    \centering
    \begin{tabular}{c|c|c|c}
    \hline
       Cosmology  &  $H_0=67$& $H_0=70$ & $H_0=74$ \\\hline
         Cosmology A & $33.6$ & $31.32$ &$28.71$ \\\hline
         Cosmology B & $2.06$ &$0.77$ &$4.46$  \\\hline
         Cosmology C, & $0.69$ &$2.21$ & $5.92$ \\\hline
    \end{tabular}
    \caption{The median of the distributions of the ratios between the computed $z_{\rm S}$ and the uncertainties provided by the Pantheon+ sample. The values for $H_0$ are in km s$^{-1}$ per Mpc.} 
    \label{tab:medianGlobal}
\end{table}

\begin{figure}
    \centering
    \includegraphics[width=0.33\hsize,height=0.3\textwidth,angle=0,clip]{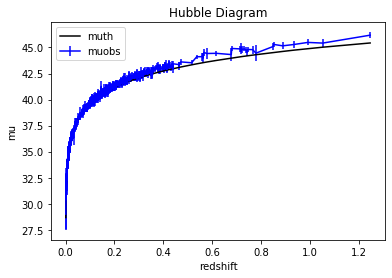}
    \includegraphics[width=0.33\hsize,height=0.3\textwidth,angle=0,clip]{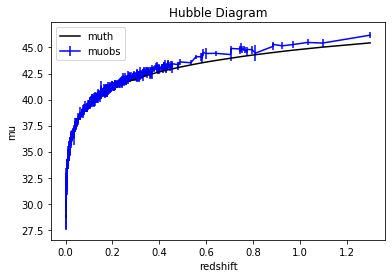}
    \includegraphics[width=0.33\hsize,height=0.3\textwidth,angle=0,clip]{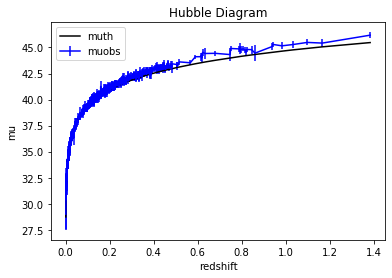}
    \includegraphics[width=0.33\hsize,height=0.3\textwidth,angle=0,clip]{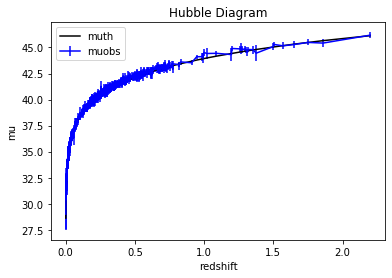}
    \includegraphics[width=0.33\hsize,height=0.3\textwidth,angle=0,clip]{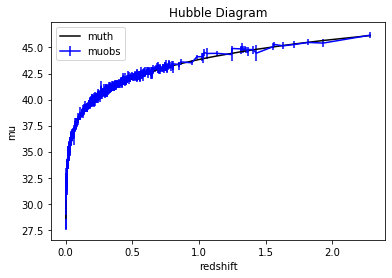}
    \includegraphics[width=0.33\hsize,height=0.3\textwidth,angle=0,clip]{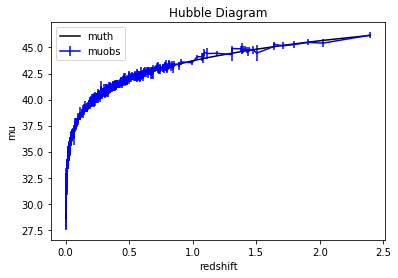}
    \includegraphics[width=0.33\hsize,height=0.3\textwidth,angle=0,clip]{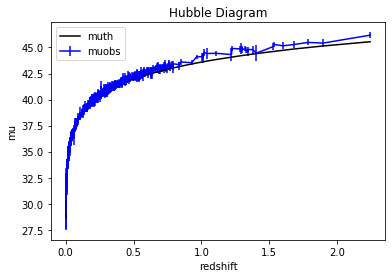}
    \includegraphics[width=0.33\hsize,height=0.3\textwidth,angle=0,clip]{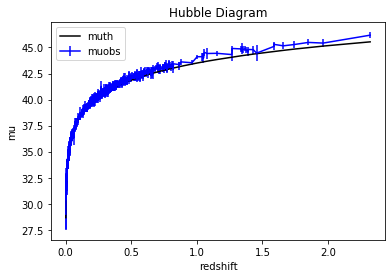}
    \includegraphics[width=0.33\hsize,height=0.3\textwidth,angle=0,clip]{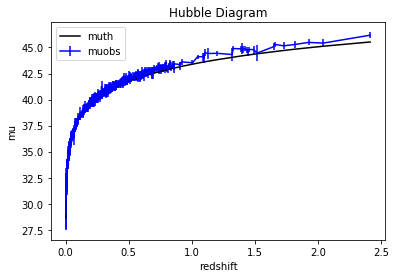}
    
    \caption{Best-fit for the Hubble Diagrams computed for our cosmological models (one for each row), using $k_1$ with data from the Pantheon+ sample, with the usual three values of $H_0$ (67, 70, 74 km s$^{-1}$ per Mpc), each for column. The black lines represent the models, while the blue curve the SNe Ia data with their respective errors. We underline that the red shift on the x-axis is the computed expansion red shift $z_{\rm C}$, which is why it changes between cosmological models.}
    \label{fig:hubblediagrams}
\end{figure}

As we did in Ref. \cite{Spallicci_2022}, we can compute the Hubble Diagrams for the Pantheon+ sample considering the $z_{\rm S}$ effects for all the Cosmology models and values of $H_0$ considered. Again, we focus on the $k_1$ parameters. The diagrams are shown in Fig. \ref{fig:hubblediagrams}. We may note that, even if we are considering a new sample, the behaviour of these diagrams is similar to the fits obtained with the Pantheon sample: the best visual fit has been reached by the Cosmology model B for all the values for $H_0$. For the other two Cosmology models, we see the theoretical black line fall below the data points at high red shifts, as was the case for the Pantheon sample. Still, for all our computations, the theoretical curve is consistent with the majority of the data points belonging to the Pantheon+ sample. We note again that the red shift on the x-axis in these plots is the cosmological red shift $z_{\rm C}$, which we note is very similar to the corresponding results obtained for the Pantheon sample.

As we did in Ref. \cite{Spallicci_2022}, to verify the goodness of our fits we have computed some statistical indicators, namely the reduced $\chi^{2}$, corresponding to the $k_{1}$ best-fit value, the root mean squared deviation (RMSD) and the normalised root mean squared deviation (NRMSD), which have been derived for all the $k_i$ parameters. The results are visualized in Table \ref{tab:statistics}. Comparing these results with the values obtained from the Pantheon computations, we note that we infer, in general, better fits: indeed, while the RMSD remains of the same order of magnitude, we note a significant decrease for the NRMSD and especially for the reduced $\chi^{2}$. We also note that, as in the Pantheon computations, the Cosmology model B reaches the best-fit results, confirming our comments in regards to Fig. \ref{fig:hubblediagrams}. The other Cosmology models show also acceptable fits, better than their older counterparts, although worse than the results obtained by Cosmology model B.

\begin{table}
    \centering
    \begin{tabular}{c|c|c|c}
    \hline
       Cosmology  &  Reduced $\chi^2$ & RMSD & NRMSD \\\hline
         Cosmology A, $H_0=67$ & 0.007 & 0.240 & 0.106 \\\hline
         Cosmology A, $H_0=70$ & 0.009 & 0.259 & 0.111 \\\hline
         Cosmology A, $H_0=74$ & 0.009 & 0.257 & 0.111\\\hline
        Cosmology B, $H_0=67$ & 0.003 & 0.186 & 0.087\\\hline
         Cosmology B, $H_0=70$ & 0.003 & 0.183 &  0.086\\\hline
         Cosmology B, $H_0=74$ & 0.003 & 0.183 & 0.086 \\\hline
         Cosmology C, $H_0=67$ & 0.006 & 0.225 & 0.104 \\\hline
        Cosmology C, $H_0=70$ & 0.006 & 0.226 & 0.104 \\\hline
          Cosmology C, $H_0=74$ & 0.006 & 0.228 & 0.104\\\hline
    \end{tabular}
    \caption{Best-fit statistics for the $k_1$ parameter in all the cases considered in our work. The NRMSD has been derived from the RMSD divided by the difference between the maximum and minimum values of the difference between the theoretical and observed distance-modulus. The mean value of the uncertainty on the observed data is $\Delta_{\mu obs}=0.243$, which is higher than the correspondent value for the Pantheon set, $\Delta_{\mu obs}=0.142$. The values for $H_0$ are in km s$^{-1}$ per Mpc.} 
    \label{tab:statistics}
\end{table}

\subsection{Adding Baryonic Acoustic Oscillations}

We now add the BAO set to our computations, as we did in Ref. \cite{Spallicci_2022}. The results related to the $k_i$ parameters are gathered in Table \ref{tab:resultskiBAO}. Comparing the results shown here with Table \ref{tab:resultskigeneral}, we note a general significant decrease in the uncertainties regarding our results, for all the considered cases. We also note that the mean values for the Cosmology model A are very similar between the two cases, while we see a significant shift to positive values for the $k_i$ parameters for the Cosmology model B, and negative ones for the Cosmology model C. We also note that we do not obtain results consistent with zero anymore for any of the considered cases. Regarding the comparison with the corresponding table found in Ref. \cite{Spallicci_2022}, we notice a negative shift of the results for the Cosmology models A and C, while a positive one for the Cosmology model B. Previously, with the combination of the Pantheon sample and BAO, we obtained for particular cases results consistent with zero, which we do not find here. This is because of the differences between the Pantheon and Pantheon+ results.

\begin{table}
    \centering
    \begin{tabular}{c|c|c|c|c}
    \hline
       Cosmology  &  $k_1$ &  $k_2$ &  $k_3$ &  $k_4$\\\hline
         Cosmology A, $H_0=67$ & $(-9.53\pm 0.03)  \times 10^{-5}$ & $(-8.84\pm 0.03)  \times 10^{-5}$ & $(-5.91\pm 0.02)  \times 10^{10}$ & $(-10.03\pm 0.04)  \times 10^{-5}$\\\hline
         Cosmology A, $H_0=70$ & $(-9.37\pm 0.03)  \times 10^{-5}$ & $(-8.71\pm 0.03)  \times 10^{-5}$ & $(-5.84\pm 0.02)  \times 10^{10}$ & $(-10.09\pm 0.03)  \times 10^{-5}$ \\\hline
         Cosmology A, $H_0=74$ & $(-9.13\pm 0.03)  \times 10^{-5}$ & $(-8.53\pm 0.03)  \times 10^{-5}$ & $(-5.71\pm 0.02)  \times 10^{10}$ & $(-9.78\pm 0.04)  \times 10^{-5}$\\\hline
         Cosmology B, $H_0=67$ & $(2.59\pm 0.05)  \times 10^{-5}$  & $(2.63\pm 0.05)  \times 10^{-5}$ & $(1.76\pm 0.04)  \times 10^{10}$ & $(2.54\pm 0.04)  \times 10^{-5}$ \\\hline
        Cosmology B, $H_0=70$ & $(5.94\pm 0.09)  \times 10^{-5}$  & $(5.39\pm 0.10)  \times 10^{-5}$ & $(2.28\pm 0.04)  \times 10^{10}$ & $(3.25\pm 0.04)  \times 10^{-5}$\\\hline
          Cosmology B, $H_0=74$ & $(4.34\pm 0.06)  \times 10^{-5}$ & $(4.49\pm 0.06)  \times 10^{-5}$ & $(3.00\pm 0.04)  \times 10^{10}$ & $(4.22\pm 0.06)  \times 10^{-5}$  \\\hline
          Cosmology C, $H_0=67$ & $(-4.80\pm 0.05)  \times 10^{-5}$ & $(-4.65\pm 0.06)  \times 10^{-5}$ & $(-3.12\pm 0.04)  \times 10^{10}$ & $(-4.91\pm 0.06)  \times 10^{-5}$ \\\hline
         Cosmology C, $H_0=70$ & $(-4.17\pm 0.07)  \times 10^{-5}$ & $(-4.08\pm 0.05)  \times 10^{-5}$ & $(-2.73\pm 0.03)  \times 10^{10}$ & $(-4.28\pm 0.06)  \times 10^{-5}$ \\\hline
         Cosmology C, $H_0=74$ & $(-3.30\pm 0.06)  \times 10^{-5}$  & $(-3.25\pm 0.06)  \times 10^{-5}$  & $(-2.16\pm 0.04)  \times 10^{10}$  & $(-3.36\pm 0.07)  \times 10^{-5}$ \\\hline
    \end{tabular}
    \caption{Results for the $k_i$ parameter considering the general best-fit for SNe Ia of the Pantheon sample together with the BAO constraints, for the three cosmological models. The values for $H_0$ are in km s$^{-1}$ per Mpc.} 
    \label{tab:resultskiBAO}
\end{table}

\section{Extended Theories of Gravity  and Extended Theories of Electromagnetism}
In this section, we discuss some results obtained considering the ETG formalism in the literature, and how we could integrate our paradigm to merge the two extensions of electromagnetism and gravity. 

We start by discussing a recent result obtained using an $f(T)$ teleparallel gravity model \cite{Benetti_2021}, in which the action is expressed by

\begin{equation}
    \mathcal{A}=\frac{1}{16 \pi G} \int d^4 x e[T+f(T)]+\mathcal{A}_m,
\end{equation}
where $f(T)$ is a generic function of the torsion scalar $T$, $\mathcal{A}_m$ is the action related to the matter terms, and $e=\sqrt{-g}$ is the metric determinant. Obtaining the Friedmann equations for this theory considering a Flat Friedmann-Robertson-Walker metric we find \cite{Benetti_2021}

\begin{equation}
    E(z)=\sqrt{\Omega_r(1+z)^4+\Omega_M(1+z)^3+\frac{1}{T_0}[f-2Tf']},
\end{equation}
where $T=6H$ holds. We note that the main difference between this equation and Eq. (\ref{E(z)}) is in the last term, which substitutes the dark energy density. This framework has been used considering three specific forms of the $f(T)$ function in concordance with real data (CMB, the Pantheon sample, BAO, DES) to derive the free parameters on which the $\Lambda$CDM model is based upon. Introducing our formalism for $z_{\rm S}$, i. e., including in computations regarding cosmological applications of ETG, like this one, also the further redshift contribution that has been analyzed in this paper, it would be interesting to consider both modifications using real data as we did in the previous sections, to understand how the two extensions compare with each other when both are employed contemporaneously to infer cosmological parameters in models without a constant dark energy contribution.

Other examples in the literature exist where ETG models have been considered to constrain observational parameters using real data. For example, \cite{Odintsov_2017} studied a particular $f(R)$-Gravity exponential model in conjunction with real data (SNe Ia, BAO, CMB, and $H(z)$) to constrain four free parameters of his theory, finding a concordance with the observed data similar to the $\Lambda$CDM-model.  Comparisons with the observations considering this exponential model were found in Refs. \cite{Yang_2010, Chen_2015}.

In \cite{carloni-etal-2019}, the dark energy has been modeled as a non-minimally coupled self-interacting fermion condensate, and this hypothesis has been tested considering three different forms for the potential associated with this scalar field considering real data, among which SNe Ia and BAO, finding results regarding their assumptions compatible with the analyzed data. This could be another example of an ETG whose results might be compared by our framework.

From a more theoretical point of view, many other possible cosmological models going beyond GR and considering ETG have been studied by the scientific community \cite{saridakis-etal-2021,Capozziello_2022}. Finding a comparison with one of these models with our approach linked to the frequency shifts, rather than on underlying gravitational and cosmological theory, would allow us to infer which of the two modifications would bring the greater impact on the cosmological results, and if or how an improvement with respect to the $\Lambda$CDM model is at hand. Of course, as already stressed in Ref.  \cite{Spallicci_2022} in regard to the non-standard frequency shift, the possibility of merging the two frameworks has itself to be confirmed by a solid theoretical background. We also note that in Ref. \cite{Spallicci_2022} a discussion regarding the inferred $z_{\rm S}$ and their relation with cosmological probes not yet considered in our analysis, like the CMB which has been widely used for the aforementioned ETG tests, is present. We here recall the important point that the CMB is generally used in conjunction with other probes for testing the dark energy hypothesis \cite{Astier-Pain-2012}, because information on this particular point depends more on the distance between the recombination epoch and us rather than on the underlying physics behind the CMB. 

\section{Discussion and conclusions}
In this paper, we have applied the methodology developed in Ref. \cite{Spallicci_2022} to the novel Pantheon+ sample, a SNe Ia set that has been used to infer the latest result regarding $H_0$ from the SHOES collaboration \cite{Riess_2022}. The aim is to derive what, assuming possible ETE effects (which may be due to: massive photon, Standard-Model Extension, and Non-Linear Electro-Magnetism theories), should the magnitude of optical shifts beyond the cosmological one be to match the observations in cosmological models without dark energy.
More specifically, we have studied three Cosmology models, A) $\Omega_{M}=0.3$, $\Omega_{K}=0$, implying a flat universe where the "cosmic triangle" relation $\Omega_{M}+\Omega_{K}+\Omega_{\Lambda}=1$, is not satisfied {\it a priori}, but {\it a posteriori}, through the effect of $z_{\rm S}$, which would act as an effective "dark energy" component; B) an open universe model, where $\Omega_{M}=0.3$ and $\Omega_{K}=0.7$, and $\Omega_{K}+\Omega_{M}=1$; C) the Einstein-de Sitter model of a flat, matter-dominated universe with $\Omega_{M}=1$. As we did in our previous work, for each model we study our results for three fixed values of $H_0$:  $67, 70, 74$ km s$^{-1}$ per Mpc.  
Given that the mock red shifts do not depend on the set considered, we move directly to the real data computations.

Comparing the results obtained by the Pantheon and Pantheon+ sets, we note that, for the individual computations, they are rather similar, with the main difference being the very high peak around $z_{\rm S}=0$ for the Pantheon+ results, probably due to the increased number of low redshift SNe Ia in the new sample. We also note a higher number of SNe Ia giving a positive contribution to the $z_{\rm S}$ computations, especially for Cosmology model B. 

More significant differences have been found for the general fits, mainly for the Cosmology Models B and C for $H_0=67$ km s$^{-1}$ per Mpc, where a shift in the sign with respect to the corresponding Pantheon computations has been noted.

Apart from the aforementioned differences, similar conclusions can be drawn from the new results. Namely, for the Cosmology A model, the derived $z_{\rm S}$ is always positive, thus corresponding to a dissipative contribution to the energy of the photons, which in turn would mean that the astrophysical objects are actually closer than what we deduce from the observed $z$. 

For the Cosmology model B, instead, we find a more heterogeneous situation regarding the $z_{\rm S}$ results: indeed, for the individual computations, we find what would seem to be an even number of positive and negative $z_{\rm S}$, especially for $H_0=67, 70$ km s$^{-1}$ per Mpc. This is reflected in the derived $k_i$ parameters, which are consistent with zero. We do not seem to recover the trend observed in the mock data for these two cases, while conclusions more similar to the Pantheon results have been found for $H_0=74$ km s$^{-1}$ per Mpc. This could be, again, because of the higher number of low redshift SNe Ia. For the general computations, we again note that the $k_i$ parameters are consistent with zero for the first two computations, confirming the individual results. This is also the reason for the changing sign of the derived $z_{\rm S}$. We recall that a negative $z_{\rm S}$ implies that the photons gain energy while crossing (inter-)galactic fields, from which we conclude that the astrophysical objects are actually farther than what we deduce from the observed $z$. 

For the Cosmology model C, regarding the individual results, we note the most similarities between the Pantheon and Pantheon+ results. More in particular, we notice a clear correlation between the observed z and the decreasing value of $z_{\rm S}$, which was also confirmed by the mock red shifts computations performed in our previous work. For the general results, we stress again the different sign for the $k_i$ parameter with respect to the Pantheon computations for $H_0=67$ km s$^{-1}$ per Mpc, which is again due to our results being consistent with $k_i=0$ for  $H_0=67, 70$ km s$^{-1}$ per Mpc. This is again because of the higher number of low redshift SNe Ia inside the Pantheon+ sample. 

We also note that for all the analyses performed in this paper, increasing the value of $H_0$ shifts the derived $z_{\rm S}$ in the same way, namely pushing them to more negative values. This is because, as we can note from Eq. (\ref{comoving flat}), $H_0$ is inversely proportional to the distance, thus, the effect of decreasing the distance is opposed by the more negative values of $z_{\rm S}$ in our comparison with the real data. We recall that this result increases the distance of the astrophysical probes from us.

As we may note also from the Pantheon findings, the main influence on our results stems from which cosmological model is considered for our computations. Namely, significant differences may be noted between the Cosmology model A with respect to the other two, especially in the magnitude and sign of the results regarding $z_{\rm S}$. This effect can be explained by the idea behind the Cosmology model A, in which the cosmic triangle relation is violated a priori, but, in some sense, recovered a posteriori, in such a way that the other cosmological parameters, not involving the Dark Energy, are fixed to the $\Lambda$CDM values. This is not true for Cosmology B and C, where the cosmological densities have been fixed in such a way that the cosmic triangle relation is respected.

Adding BAO has the same effects noted for the Pantheon computations, namely shifting the results towards negative values of the $k_i$ parameters for Cosmology models A and C, and towards positive ones for Cosmology model B. This was expected because we did not change the set of BAO used in our computations.

The Pantheon+ set presents also the error on the observed red shifts, allowing us a comparison with the inferred $z_{\rm S}$. We noted that the non-standard shifts are, in the majority of the cases, bigger than the provided errors, thus, we can conclude that they cannot be encompassed by uncertainties in the observations. This was one of the proposed perspectives in the previous paper on this framework. A comparison between ETE and ETG is in order, to understand which of these two extensions could bring the highest contribution to the fit of the observed results.

\section*{Data availability statement}
The data analyzed in this work are available at https://github.com/PantheonPlusSH0ES/DataRelease for the Pantheon+ sample \cite{Scolnic_2022}, and in \cite{beutler-etal-2011,blake-etal-2011,ross-etal-2015, dumasdesbourboux-etal-2020,alam-etal-2021} for the BAO set.

\section*{Acknowledgements}
.
Acknowledgements are due to J.A. Helay\"el-Neto (Rio de Janeiro) for our common work on the Extended Theories of Electromagnetism, to S. Savastano (Potsdam) for the work on the Cobaya routine, and to M. Lòpez Corredoira (La Laguna) for general comments and inputs. GS is
grateful to the LPC2E laboratory for its hospitality during the work
period on this manuscript. GS and SC acknowledge the support of Istituto Nazionale di Fisica Nucleare, Sez. di Napoli, Iniziative
Specifiche MOONLIGHT-2 and QGSKY.

\bibliography{references_spallicci_220129}

\end{document}